\documentstyle[12pt,epsf]{article}

\catcode`@=11
\def\chkspace{%
  \relax   
  \begingroup\ifhmode\aftergroup\dochksp@ce\fi\endgroup}
\def\dochksp@ce{%
  \unskip              
  \futurelet\chkspct@k\d@chkspc  
}
\def\d@chkspc{%
  \let\nxtsp@ce=\relax
  \ifx\chkspct@k.\else     
    \ifx\chkspct@k,\else
      \ifx\chkspct@k;\else
        \ifx\chkspct@k!\else
          \ifx\chkspct@k?\else
            \ifx\chkspct@k:\else
              \ifx\chkspct@k)\else
              \ifx\chkspct@k(\else
                \ifx\chkspct@k]\else
                  \ifx\chkspct@k-\else
                    \ifx\chkspct@k\egroup\else  
                      \let\nxtsp@ce=\put@space  
                    \fi
                  \fi
                \fi
              \fi
              \fi
            \fi
          \fi
        \fi
      \fi
    \fi
  \fi
  \nxtsp@ce
}
\def\put@space{$\;$}
\catcode`@=12

\def\ra{{$\rightarrow$}\chkspace}

\def\etal{{\it et al.}\chkspace}

\def\adhoc{{\it ad hoc}\chkspace}

\def\ep{{e$^+$e$^-$}\chkspace}

\def\gluino{\relax\ifmmode \tilde{g} \else $\tilde{g}$ \fi\chkspace}

\chkspace
\chkspace
\def\m0{$M_{0}$}\chkspace
\def\m0m{$M_{0}MAX$}\chkspace
\chkspace
\chkspace

\def\qq{q$\overline{\rm q}$\chkspace}

\def\bbrm{\relax\ifmmode {\rm b}\bar{\rm b}
       \else ${\rm b}\bar{\rm b}$ \fi\chkspace}
\def\bb{$b\bar{b}$ \chkspace}
\def\ccrm{\relax\ifmmode {\rm c}\bar{\rm c}
       \else ${\rm c}\bar{\rm c}$ \fi\chkspace}

\def\tt{\relax\ifmmode {\rm t}\bar{\rm t}
       \else ${\rm t}\bar{\rm t}$ \fi\chkspace}
\def\ss{\relax\ifmmode {\rm s}\bar{\rm s}
       \else ${\rm s}\bar{\rm s}$ \fi\chkspace}
\def\uu{\relax\ifmmode {\rm u}\bar{\rm u}
       \else ${\rm u}\bar{\rm u}$ \fi\chkspace}
\def\dd{\relax\ifmmode {\rm d}\bar{\rm d}
       \else ${\rm d}\bar{\rm d}$ \fi\chkspace}

\def\qqg{\relax\ifmmode {\rm q}\overline{\rm q}{\rm g}
\else q$\overline{\rm q}$g \fi\chkspace}

\def\bbg{$b\overline{b}g$\chkspace}

\def\qqgg{{q$\overline{\rm q}$gg}\chkspace}

\def\afb{\relax\ifmmode A_{FB} \else
{{$A_{FB}$}}\fi\chkspace}
\def\afbb{\relax\ifmmode A_{FB}^b \else
{{$A_{FB}^b$}}\fi\chkspace}
\def\pafb{\relax\ifmmode \tilde{A}_{FB} \else
{{$\tilde{A}_{FB}$}}\fi\chkspace}
\def\pafbb{\relax\ifmmode \tilde{A}_{FB}^b \else
{{$\tilde{A}_{FB}^b$}}\fi\chkspace}

\def\pafbzo{\relax\ifmmode \tilde{A}_{FB}|_{O(0)} \else
{{$\tilde{A}_{FB}|_{O(0)}$}}\fi\chkspace}
\def\pafbfo{\relax\ifmmode \tilde{A}_{FB}|_{\oalp} \else
{{$\tilde{A}_{FB}|_{\oalp}$}}\fi\chkspace}
\def\pafbso{\relax\ifmmode \tilde{A}_{FB}|_{\oalpsq} \else
{{$\tilde{A}_{FB}|_{\oalpsq}$}}\fi\chkspace}
\def\pafbto{\relax\ifmmode \tilde{A}_{FB}|_{\oalpc} \else
{{$\tilde{A}_{FB}|_{\oalpc}$}}\fi\chkspace}

\def\pafbbzo{\relax\ifmmode \tilde{A}_{FB}^b|_{O(0)} \else
{{$\tilde{A}_{FB}^b|_{O(0)}$}}\fi\chkspace}
\def\pafbbfo{\relax\ifmmode \tilde{A}_{FB}^b|_{\oalp} \else
{{$\tilde{A}_{FB}^b|_{\oalp}$}}\fi\chkspace}
\def\pafbbso{\relax\ifmmode \tilde{A}_{FB}^b|_{\oalpsq} \else
{{$\tilde{A}_{FB}^b|_{\oalpsq}$}}\fi\chkspace}
\def\pafbbto{\relax\ifmmode \tilde{A}_{FB}^b|_{\oalpc} \else
{{$\tilde{A}_{FB}^b|_{\oalpc}$}}\fi\chkspace}

\def\afbo0{\tilde{A}_{FB}|_{O(0)}}
\def\afbo1{\tilde{A}_{FB}|_{\oalp}}
\def\afbo2{\tilde{A}_{FB}|_{\oalpsq}}
\def\afbo3{\tilde{A}_{FB}|_{\oalpc}}

\def\lam{\relax\ifmmode \Lambda_{\overline{MS}}
       \else {{$\Lambda_{\overline{MS}}$}}\fi\chkspace}
\def\lamuds{\relax\ifmmode \Lambda^{(3)}_{\overline{MS}}
       \else {{$\Lambda^{(3)}_{\overline{MS}}$}}\fi\chkspace}
\def\lamudsc{\relax\ifmmode \Lambda^{(4)}_{\overline{MS}}
       \else $\Lambda^{(4)}_{\overline{MS}}$\fi\chkspace}
\def\lamudscb{\relax\ifmmode \Lambda^{(5)}_{\overline{MS}}
       \else $\Lambda^{(5)}_{\overline{MS}}$\fi\chkspace}

\def\alp{\relax\ifmmode \alpha_s\else $\alpha_s$\fi\chkspace}
\def\alpbar{\relax\ifmmode \bar{\alpha_s}
       \else $\bar{\alpha_s}$\fi\chkspace}
\def\alpmz{\relax\ifmmode \alpha_s(M_Z)\else $\alpha_s(M_Z)$\fi\chkspace}
\def\alpmzsq{\relax\ifmmode \alpha_s(M_Z^2)
       \else $\alpha_s(M_Z^2)$\fi\chkspace}

\def\oalp{\relax\ifmmode O(\alpha_s)\else{{O($\alpha_s$)}}\fi\chkspace}
\def\oalpsq{\relax\ifmmode O(\alpha_s^2)
           \else{{O($\alpha_s^2$)}}\fi\chkspace}
\def\oalpc{\relax\ifmmode O(\alpha_s^3)
           \else{{O($\alpha_s^3$)}}\fi\chkspace}
\def\oalpf{\relax\ifmmode O(\alpha_s^4)
           \else{{O($\alpha_s^4$)}}\fi\chkspace}

\def\rb{\relax\ifmmode R_3^b/R_3^{all}
           \else{{$R_3^b/R_3^{all}$}}\fi\chkspace}
\def\rc{\relax\ifmmode R_3^c/R_3^{all}
           \else{{$R_3^c/R_3^{all}$}}\fi\chkspace}
\def\ruds{\relax\ifmmode R_3^{uds}/R_3^{all}
           \else{{$R_3^{uds}/R_3^{all}$}}\fi\chkspace}
\def\ri{\relax\ifmmode R_3^i/R_3^{all}
           \else{{$R_3^i/R_3^{all}$}}\fi\chkspace}
\def\rj{\relax\ifmmode R_3^j/R_3^{all}
           \else{{$R_3^j/R_3^{all}$}}\fi\chkspace}
\def\alpi{\relax\ifmmode \alpha^i_s/\alpha^{all}_s
           \else{{$\alpha^i_s/\alpha^{all}_s$}}\fi\chkspace}

\def\z0{{$Z^0$}\chkspace}
\def\Dst{\relax\ifmmode {\rm D}^* \else {D$^*$}\fi\chkspace}
\def\Dpl{\relax\ifmmode {\rm D}^+ \else {D$^+$}\fi\chkspace}
\def\D0{\relax\ifmmode {\rm D}^0 \else {D$^0$}\fi\chkspace}
\def\Kst{\relax\ifmmode {\rm K}^* \else {K$^*$}\fi\chkspace}
\def\K0{\relax\ifmmode {\rm K}^0_s \else {K$^0_s$}\fi\chkspace}
\def\Kpl{\relax\ifmmode {\rm K}^+ \else {K$^+$}\fi\chkspace}
\def\Kstz{\relax\ifmmode {\rm K}^{*0} \else {K$^{*0}$}\fi\chkspace}

\def\ZPC{{\em Z. Phys.} C}

\def\ra{{$\rightarrow$}\chkspace}

\def\be{\begin{equation}}
\def\ee{\end{equation}}
\def\bea{\begin{eqnarray}}
\def\eea{\end{eqnarray}}

\def\ep{{$e^+e^-$}\chkspace}
\def\bbg{{$b\bar{b}g$}\chkspace}
\def\gbb{{$g \rightarrow b\bar{b}$}\chkspace}
\def\bbgg{{$b\bar{b}gg$}\chkspace}
\def\qqgg{{$q\bar{q}gg$}\chkspace}
\def\z0{$Z^0$}
\def\xb{{$x_B$}\chkspace}
\def\bb{{$b\bar{b}$}\chkspace}

\def\rate{{$g_{b\bar{b}}$}\chkspace}
\def\etal{{\it et al.}\chkspace}
\def\adhoc{{\it ad hoc}\chkspace}
\def\gcc{{$g\rightarrow c\bar{c}$}\chkspace}
\def\qqg{{$q\bar{q}g$}\chkspace}
\def\gqq{{$g\rightarrow q\bar{q}$}\chkspace}
\def\qqbb{{$q\bar{q}b\bar{b}$}\chkspace}
\def\qqcc{{$q\bar{q}c\bar{c}$}\chkspace}
\def\qq{{$q\bar{q}$}\chkspace}

\renewcommand{\baselinestretch}{1.}

 1

\catcode`\@=11 
\makeatletter
\def\@seccntformat#1{\csname the#1\endcsname.\hskip 1em}

\makeatother
\begin{document}
\thispagestyle{empty}
\begin{flushright}
{\footnotesize\renewcommand{\baselinestretch}{.75}
  SLAC--PUB--8150\\
\vspace{0.1cm}
May 1999\\
}
\end{flushright}

\vskip 0.5truecm
 
\begin{center}
 {\Large \bf Tests of QCD using Heavy Flavors at SLD$^*$}
 
\vspace {1.4cm}
  {\bf Danning Dong}
\vspace{0.3cm}\\
Massachusetts Institute of Technology, Cambridge, MA 02139\\
\vspace{0.5cm}
{\it Representing} \\
\vspace{0.3cm}
{\bf The SLD Collaboration$^{**}$}\\
\vspace{0.2cm}
Stanford Linear Accelerator Center\\
Stanford University, Stanford, CA 94309
\end{center}
 
\normalsize
 
\vspace{1cm} 
\begin{center}
{\bf Abstract }
\end{center}

{
\linewidth = 100mm
\indent
We present preliminary results on three SLD analyses: 
the gluon energy spectrum in 3-jet \bbg events,
the rate of \gbb, and the $b$ fragmentation function in \z0 decays.
The gluon energy spectrum, measured over the 
full kinematic range, is compared 
with perturbative QCD predictions.  We 
set new 95\% C.L.\ limits on the anomalous 
chromomagnetic coupling of the $b$ quark: $-0.09<\kappa<0.06$.
\rate is measured to be 
(3.07 $\pm$ 0.71 (stat) $\pm$ 0.66 (syst)) $\times 10^{-3}$.
The inclusive $B$ hadron energy distribution in \z0 decays is measured for the 
first time over the full kinematic range, using a novel $B$ hadron energy 
reconstruction technique.  
Several models of $b$ fragmentation including JETSET $+$ Peterson
are excluded by the data.  The average scaled $B$ hadron energy of 
the weakly decaying $B$ hadron is measured to be 
\xb $=$ 0.713 $\pm$ 0.005 (stat) $\pm$ 0.007 (syst) 
$\pm$ 0.002 (model).  All three 
measurements take advantage of the small and stable SLC interaction 
point as well as the excellent vertexing and tracking capabilities of the 
upgraded CCD-pixel vertex detector. 
}

\vfil
 
\noindent
\begin{center}
{\it Invited talk presented at the XXXIVth Rencontres de Moriond\\
QCD and High Energy Hadronic Interactions\\ 
Les Arcs, Savoie, France\\
March 20-27, 1999}
\end{center}
\vskip .3truecm
\noindent
$^*$ Work supported in part by Department of Energy contracts
DE-FC02-94ER40818 and DE-AC03-76SF00515.
%
%
\eject
  
\section{Introduction}
\noindent 
One of the powerful aspects of QCD is that it provides a framework 
to understand the production of hadronic jets.  
Perturbative QCD (pQCD) in principle allows us 
to calculate infrared finite quantities as an expansion of the strong 
coupling $\alpha_s$ and its predictions are subject to 
precision experimental tests.  
The nonperturbative fragmentation process, 
which transforms quarks and gluons into hadrons, however, 
has not been well-understood theoretically, nor has it been
well-measured experimentally.  
This limits the precision in our understanding of 
the production and structure of hadronic jets and 
will likely affect the precision of physics studies at future colliders,
where large numbers of hadrons will be produced.
In view of this, it is important both to test pQCD 
and to precisely measure quantities 
that will help extract the nonperturbative effects in the 
fragmentation process.  Heavy flavors are especially well-suited for 
this purpose.  The mass of the $b$ quark provides a natural 
cutoff in pQCD calculations, a feature that allows
a number of precise pQCD predictions to be made.
Furthermore, \ep \ra \z0 environment is a particularly clean.  
When a $B$ hadron is highly boosted, 
its decay vertex is significantly displaced from the interaction 
point (IP), allowing 
high efficiency and purity $B$ tagging.

Here we report 
three SLD~\cite{sld} analyses involving heavy quark production.
SLD's strength to study $B$ physics largely resides in 
the upgraded CCD-pixel vertex detector (VXD3)~\cite{vxd3}.   Using VXD3, the 
 impact parameter resolution is 11$\mu$m (23$\mu$m) transverse to 
(along) the beam axis for high momentum tracks. 
SLC's small and stable IP 
can be measured to a resolution of about 4.4$\mu$m in the plane 
transverse to the beam axis.  This allows not only the selection of 
a high purity $B$ sample, but also the precise measurement of 
kinematic quantities.

\section{\bf The Gluon Energy Spectrum in \z0 \ra \bbg }

In studying \ep \ra 3-jet events , the key is to know the origin of 
each jet.  In previous studies, jets were usually energy-ordered and 
the lowest is assumed to be the gluon jet.  This is correct about 80\% of the
time.  However, the bias towards selecting low energy gluon jets 
makes this method undesirable for testing pQCD predictions 
of the gluon jet energy spectrum, to which 
high energy gluon jets are most sensitive.  

In this analysis~\cite{bbg}, the gluon jet is 
tagged explicitly by identifying the two jets that contain $B$ hadrons. 
A 3-jet event (selected using the JADE algorithm at $y_{cut}=0.02$) 
is tagged as a \bbg event if exactly two of the three jets 
each contain two or more tracks with normalized transverse impact 
parameter $d/\sigma_d > 3$.  The remaining jet, which does not satisfy
this criterion, is tagged as the gluon jet.  
In 2.5\% (12.5\%) of the selected \bbg events, 
the (second) highest energy jet is tagged as the gluon jet, 
covering the full kinematic range.
Non-\bb events (4\%), non-\bbg events (21\%), and \bbg events
in which the gluon jet is wrongly tagged (1\%) are treated as backgrounds and
are subtracted from the data.  The resulting distribution of the 
scaled gluon energy 
$x_g = 2E_g/E_{CM}$ is corrected for effects of efficiency and resolution.
The fully corrected spectrum is shown in Figure 1.  The cutoff
 at low $x_g$ is caused by the finite $y_{cut}$ value used
 in jetfinding. Leading order (LO) and next-to-leading order (NLO) 
QCD predictions~\cite{NLO} 
describe the general but not the
detailed behavior of the spectrum.  Higher order QCD effects
are clearly important. Parton shower Monte Carlo is able to describe
the data.

\parbox[b]{2.7 in}{
  \epsfxsize 2.7 in
  \epsffile{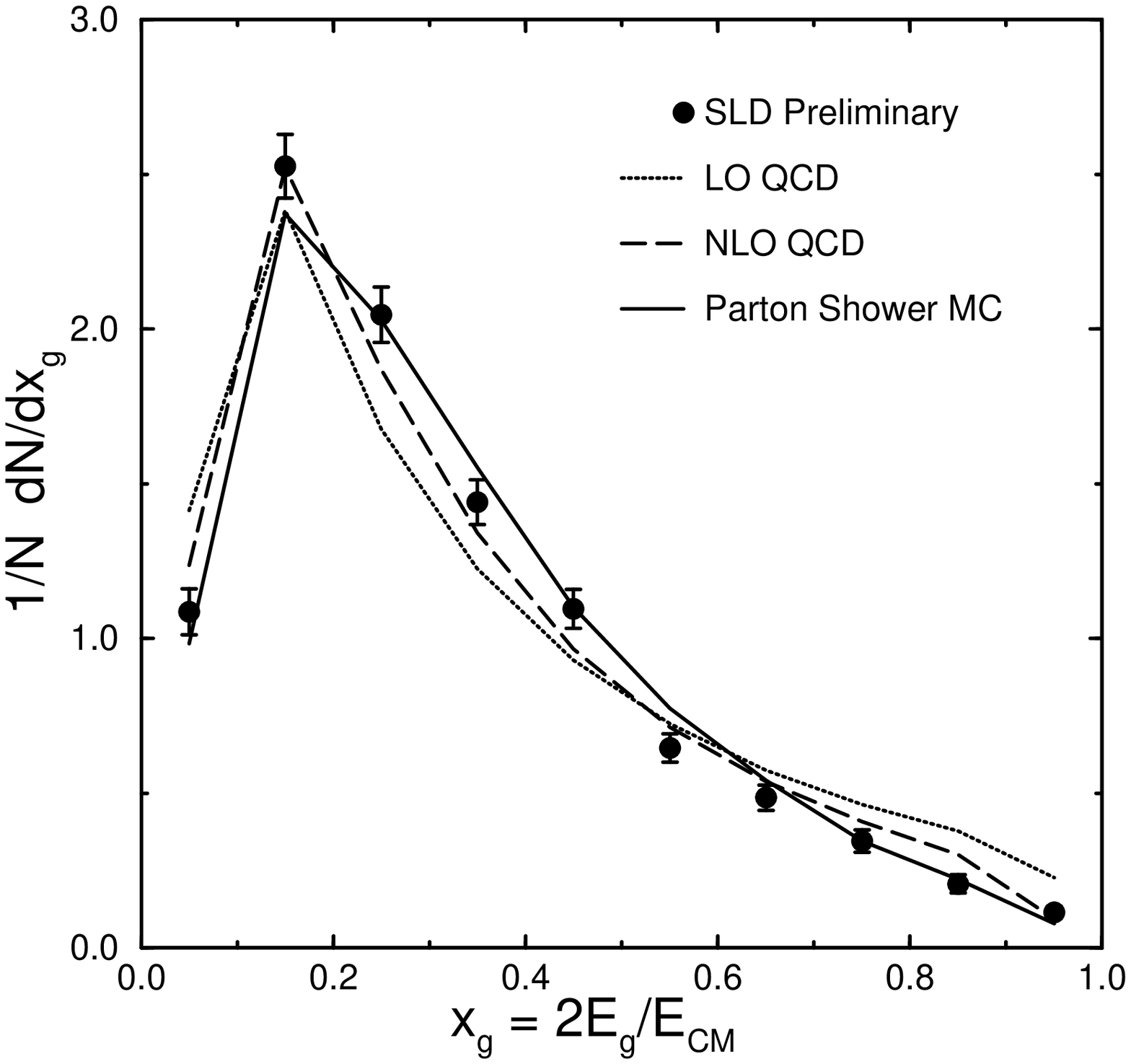}
  { \small Figure 1. The measured scaled gluon energy spectrum. }
}
\ \hspace{0.1cm} \ 
\vspace{0.1cm}
\parbox[b]{3.15in} { \vspace{0.2cm}
\hspace{1.3em}
The gluon energy spectrum is particularly sensitive to 
the presence of an anomalous chromomagnetic coupling of 
the $b$ quark in the QCD Lagrangian~\cite{rizzo},  
\begin{equation}
{\cal L}^{b\bar{b}g} =  g_s\bar{b}T_a \{ \gamma_{\mu} + 
\frac{i\sigma_{\mu\nu}k^{\nu}}{2m_b}(\kappa - i \tilde{\kappa}\gamma_5)\} 
bG_a^{\mu}
\end{equation}
\noindent
where $\kappa$ and $\tilde{\kappa}$ parameterize the anomalous 
chromomagnetic and chromoelectric moments, 
respectively, which might arise from physics beyond the SM. 
Setting $\tilde{\kappa}$ to zero, a fit of the theoretical prediction 
to the data yields $\kappa =$ $-0.02$ $\pm$ 0.04 (preliminary), which
is consistent with zero.  The 95\% C.L.\ limits are 
$-0.09<\kappa<0.06$ (preliminary).
}

\noindent 
\section{\bf The Rate of Secondary \bb Production via \gbb}
The process of gluon splitting into a quark-antiquark pair is poorly known,
both theoretically and experimentally, despite the fact that this is one
of the elementary processes in QCD.  The rate of \gbb, 
\rate, defined as the fraction of hadronic events in which a gluon 
splits into a \bb pair, is an infrared finite quantity and can be
calculated using pQCD because the mass of the $b$ quark provides a 
natural cutoff~\cite{sm}.  However, \rate is sensitive to both 
the $\Lambda^5_{\overline{MS}}$ parameter and
the $b$ quark mass, which results in a substantial theoretical uncertainty
in the calculation of \rate. 
The limited accuracy of the \gbb prediction is one

\parbox[b]{2.7 in}{
  \epsfxsize 2.6 in
  \epsffile{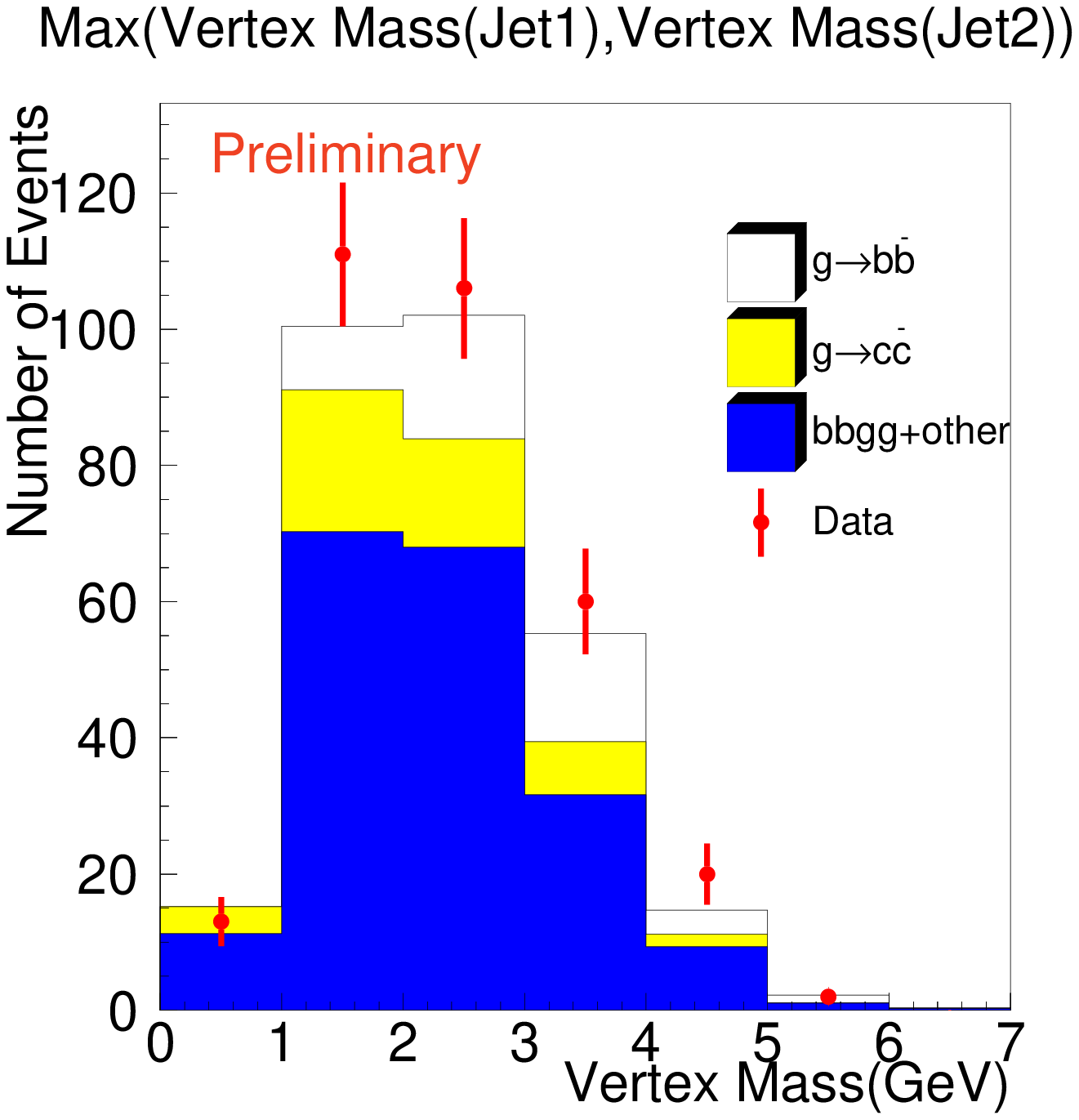}
  { \small 
Figure 2. The larger of the two vertex masses.
}
}
\ \hspace{0.1cm} \ 
\vspace{0.1cm}
\parbox[b]{3.15in} { 
\vspace{0.1cm}
 of the
 main sources of uncertainty in the measurement
of the partial decay width of the \z0 \ra \bb.  In addition,
about 50\% of the $B$ hadrons are produced via gluon splitting at Tevatron,
and an even larger fraction is expected at the LHC.  A better knowledge
of this process will improve the theoretical modeling and predictions
of heavy flavor production at such colliders.

\hspace{1.3em}
Since the background in this analysis is very large, 
the main task is to enhance the signal to background ratio. 
The 4-parton final state of the signal, \qqbb, suggests 
that we select 4-jet hadronic events. We used 
the Durham algorithm with $y_{cut}=0.008$. 
Most \z0 \ra \qq events and \z0 \ra \qqg events in which the gluon 
does not split are rejected,
}
 but the majority of 
the selected events are still backgrounds
from \qqgg and \qqcc. 
 Background of the type \qqgg 
where $q \neq b$ or $c$ can be mostly removed 
by looking for heavy hadrons in the selected events. 
Using a topological vertexing algorithm~\cite{zvnim}, 
secondary vertices are searched for in the two jets forming the smallest 
angle. About 300 events in which both jets contain 
a secondary vertex are selected.  
Figure 2 shows the distribution of the
larger of the two vertex masses, together with the breakdown of Monte Carlo 
events into the signal and the backgrounds.  

We then focus on distinguishing the topology of \gbb events
from that of \bbgg events.  In many events, the two selected jets
actually originate from the same $b$ jet.  Figure 3 shows the
angular separation between the two vertices.  
The high IP and vertex resolutions give us a good
discriminating power between the signal and the background, about
half
\parbox[b]{2.8in}{
  \epsfxsize 2.8 in
  \epsfysize 1.4 in
  \epsffile{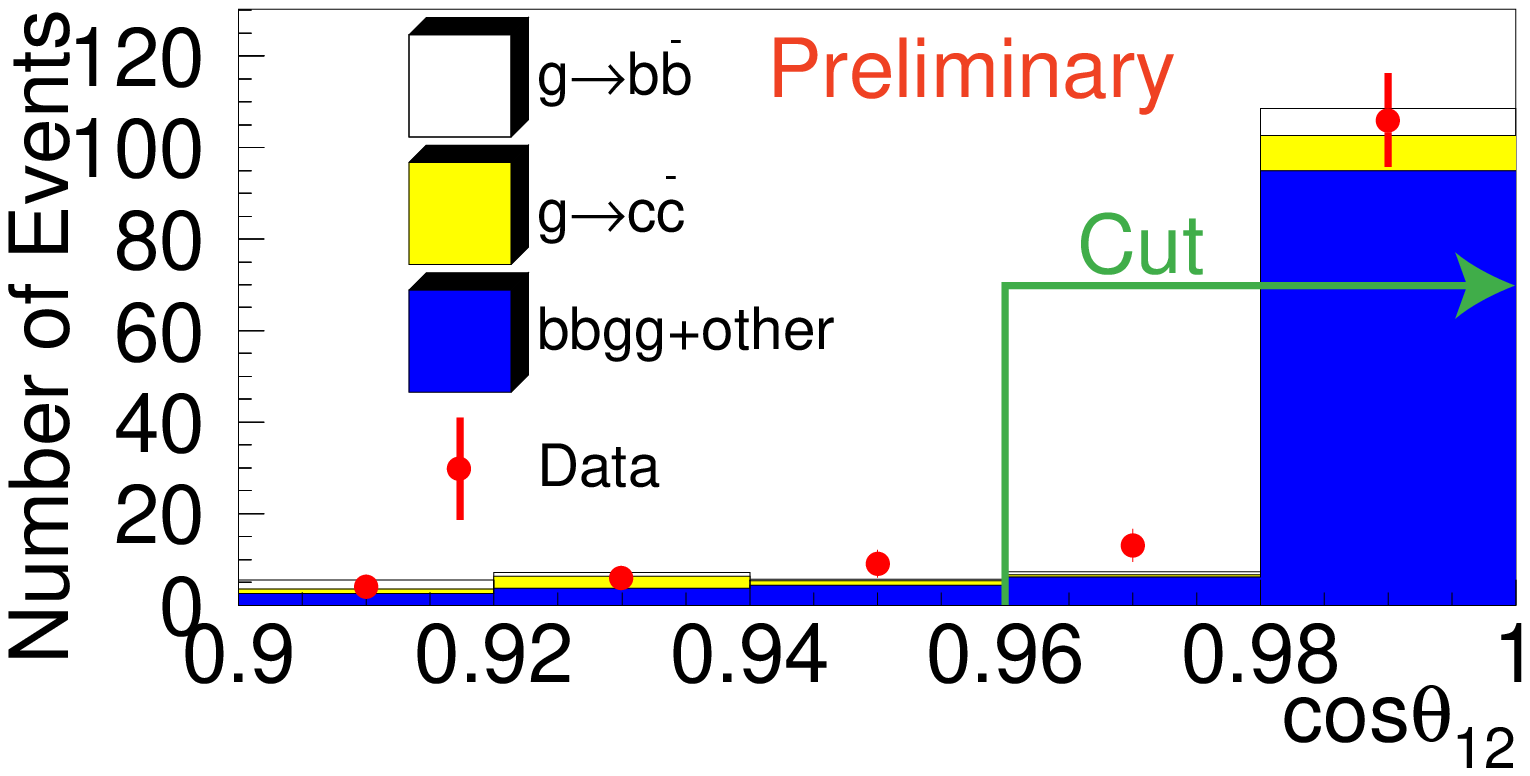}
  {\small Figure 3.
Distribution of the cosine of 
the angle between the two jets 
forming the smallest angle, 
for data (points) and in Monte Carlo (histogram).}
}
\ \hspace{0.1cm} \ 
\vspace{0.1cm}
\parbox[b]{3.3in} { 
\vspace{0.1cm}
 of which peak at cos$\theta \sim 1$. In order 
to enhance the signal to background ratio, we require 
$-$0.2 $<$ cos$\theta <$ 0.96.  Since the radiated
virtual gluon in \z0 \ra \qqg is polarized in the plane of the 
three partons, the subsequent gluon splitting favors \gqq 
emission out of this plane.  We therefore consider 
a variable similar to the Bengtsson-Zerwas angle, which
is the angle between the plane formed by the two jets containing
secondary vertices and the plane formed by the other two jets.
Figure 4 shows the distribution of $\mid$ cos$\alpha \mid$.
Events with
}
\noindent
\parbox[b]{2.8in}{
  \epsfxsize 2.8 in
  \epsfysize 1.6 in
  \epsffile{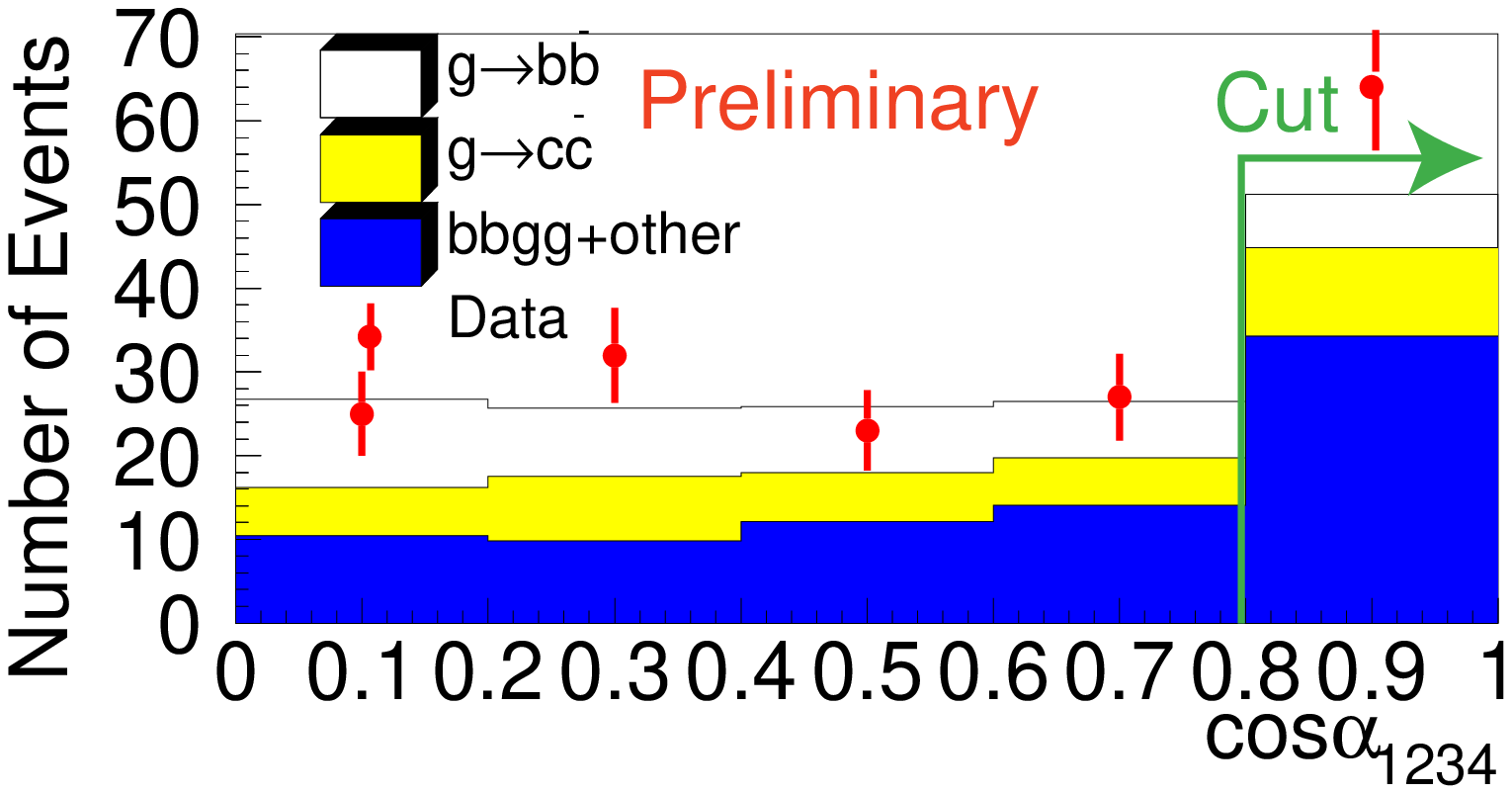}
  {\small Figure 4.
Distribution of the cosine of 
the angle between the two planes, 
for data (points) and in Monte Carlo (histogram).
\vspace{0.4cm}
}
}
\ \hspace{0.1cm} \ 
\vspace{0.2cm}
\parbox[b]{3.3in} { \vspace{-0.03cm}
  $\mid$ cos$\alpha\mid >$ 0.8 are rejected.
Finally, since $b$ jets from the gluon splitting tend to be 
softer than 
other jets, we require the total energy of the two selected jets
to be less than 36 GeV.
After these cuts, the background consists of 49\% \bbgg, 34\% \gcc
and the remaining 17\% \qqgg, where $q \neq b$ or $c$.  To further 
reduce the \gcc background, we require the maximum vertex mass 
to be greater than  2.0 GeV.
62 events in the data survived all these selection cuts.  We subtract
the expected background of 27.6$\pm$1.2 events and divide the 
selection efficiency of 3.86\% to obtain 
the preliminary measured value of \rate 
} 
\begin{equation}
g_{b\bar{b}} = (3.07\pm0.71(stat)\pm0.66(syst))\times 10^{-3}.
\end{equation}

\section{The b Quark Fragmentation Function}
According to the factorization theorem, the heavy quark fragmentation 
function can be described as a convolution of perturbative
and non-perturbative effects.  For the $b$ quark, 
the perturbative calculation is in principle understood~\cite{MN,jaffe}.
Nonperturbative effects have been parametrized in 
both model-dependent~\cite{kart,bowler,pete,lund,collins,MN} and 
model-independent approaches~\cite{jaffe,lisa,BCFY}.

It is experimentally challenging 
to measure the $b$ quark fragmentation function to a level of 
precision sufficient to distinguish among the various models.
Since the $b$ quark fragmentation function is the probability 
distribution of the fraction of the momentum of the $b$ quark 
carried by the $B$ hadron, the most sensitive experimental determination 
of the shape of the 
$b$ fragmentation function is expected to come 
from a precise direct measurement of the 
$B$ hadron energy (or momentum) distribution.  The difficulty of such 
a measurement stems mostly from the fact that most of the $B$ decays 
can only be partially reconstructed, 
causing a fraction of the $B$ 
energy to be missing from the $B$ decay vertex.
Recent direct measurements at LEP~\cite{delphi93,aleph95} 
and SLD~\cite{sld96} have used overall energy-momentum constraints and 
calorimetric information to extract this missing energy in a 
sample of semi-leptonic $B$ decays.  These 
measurements suffer from low 
statistics as well as poor $B$ energy resolution at low energy,
and hence have a relatively weak discriminating power between different shapes 
of the fragmentation function.
Indirect measurements~\cite{early} such as the measurement of the 
lepton spectrum 
have been used to constrain the average $B$ energy.  These measurements,
however, are not sensitive to the shape of the $B$ hadron energy distribution.  

Here we describe a new method for reconstructing individual 
$B$ hadron energy with good resolution over the full kinematic
range while achieving a much higher efficiency~\cite{ichep98}.
We reconstruct secondary $B$ decay vertices inclusively with high 
efficiency using a topological vertexing algorithm~\cite{zvnim}.
At SLD, the $B$ flight direction, pointing along the line joining
the IP and the secondary vertex, is well-measured 
because of the very small IP error and the excellent vertex resolution.
Therefore the transverse momentum $P_t$ of tracks associated with 
the vertex relative to the $B$ flight direction is also well-measured.  
We then obtain the invariant mass $M_{ch}$ and the total energy $E_{ch}$ of the
associated tracks, assigning the pion mass to each charged track.  Furthermore,
we assume that the true mass of the $B$ hadron decayed at the vertex, $M_B$,
is equal to the $B^0$ meson mass.  An upper limit 
on the mass of the missing particles is then found to be
$M_{0max}^2 = M_B^2 - 2M_B\sqrt{M_{ch}^2 + P_t^2} + M_{ch}^2. $
Since the true missing mass $M_0^{true}$ is 
often rather close to $M_{0max}$,
$M_{0max}$ is subsequently used as an 
estimate of $M_0^{true}$ 
to solve
 for the longitudinal momentum of the missing particles from kinematics,
and hence the energy of missing particles $E_0$.  The $B$ hadron
energy is then $E_B = E_{ch} + E_0$.  
Since $ 0 \leq  M_0^{true} \leq M_{0max}$, 
the $B$ energy is well-constrained when $M_{0max}$ is small.  In addition,
most $uds$ and $c$ backgrounds are concentrated at large
$M_{0max}$.  
We choose an {\it ad hoc} upper cut on the $M_{0max}^2$
to achieve a nearly $x_B$-independent $B$ selection efficiency
of about 3.9\%; the estimated purity is about 99.5\%.  
A total of 1920 vertices in the 1996-97 data satisfy all selection cuts.
Figure 5 shows the distribution of the reconstructed scaled weakly
decaying $B$ hadron energy for data and Monte Carlo.

We examine the normalized difference between the true and reconstructed 
$B$ hadron energies for Monte Carlo events.  The distribution is fitted 
by a double Gaussian, resulting in a core width 
(the width of the narrower Gaussian) of 10.4\% and a tail width 
(the width of the wider Gaussian) of 23.6\% with a core fraction of 83\%.  
The core width is essentially independent of $x_B$, another feature that
makes this method unique.

After background subtraction, 
the distribution of the reconstructed scaled $B$ hadron energy  
is compared with a set of {\em ad hoc} functional forms of the 
$x_B$ distribution in order to estimate the variation 
in the shape of the $x_B$ distribution. 
For each functional form, the default SLD Monte Carlo is re-weighted 
and then compared with the data bin-by-bin and a $\chi^2$ is computed.  
The minimum $\chi^2$ is found by varying the input parameter(s).
The Peterson function~\cite{pete}, two \adhoc generalizations of the 
Peterson function~\cite{aleph95}, 
and an 8th-order polynomial are consistent with the data.  
We exclude the functional forms proposed by 
BCFY~\cite{BCFY}, Collins and Spiller~\cite{collins}, 
Kartvelishvili~\cite{kart}, Lund~\cite{lund} and 
a power function of the form $f(x)=x^\alpha(1-x)^\beta$.

\parbox[b]{2.77 in}{
  \vspace{-0.4cm}
  \vspace{0.2cm}
  \epsfxsize 2.77 in
  \epsffile{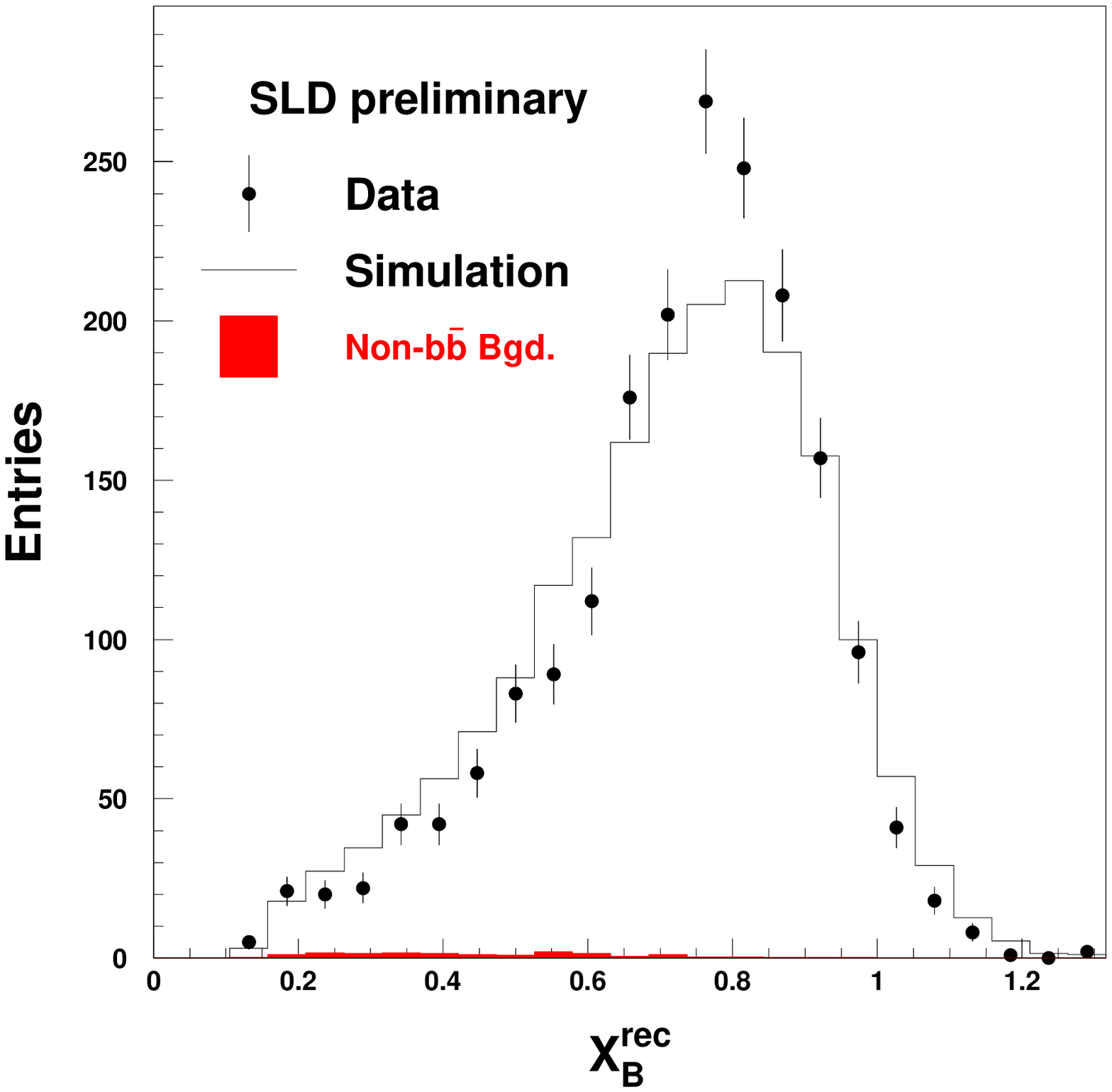}
{\small Figure 5.
Distribution of the reconstructed scaled $B$ hadron energy.
\vspace{0.17cm}
}
}
\ \hspace{0.2cm} \ 
\vspace{0.1cm}
\parbox[b]{2.8 in}{
  \vspace{.5cm} 
  \epsfxsize 2.64 in
  \epsfysize 2.5 in
  \epsffile{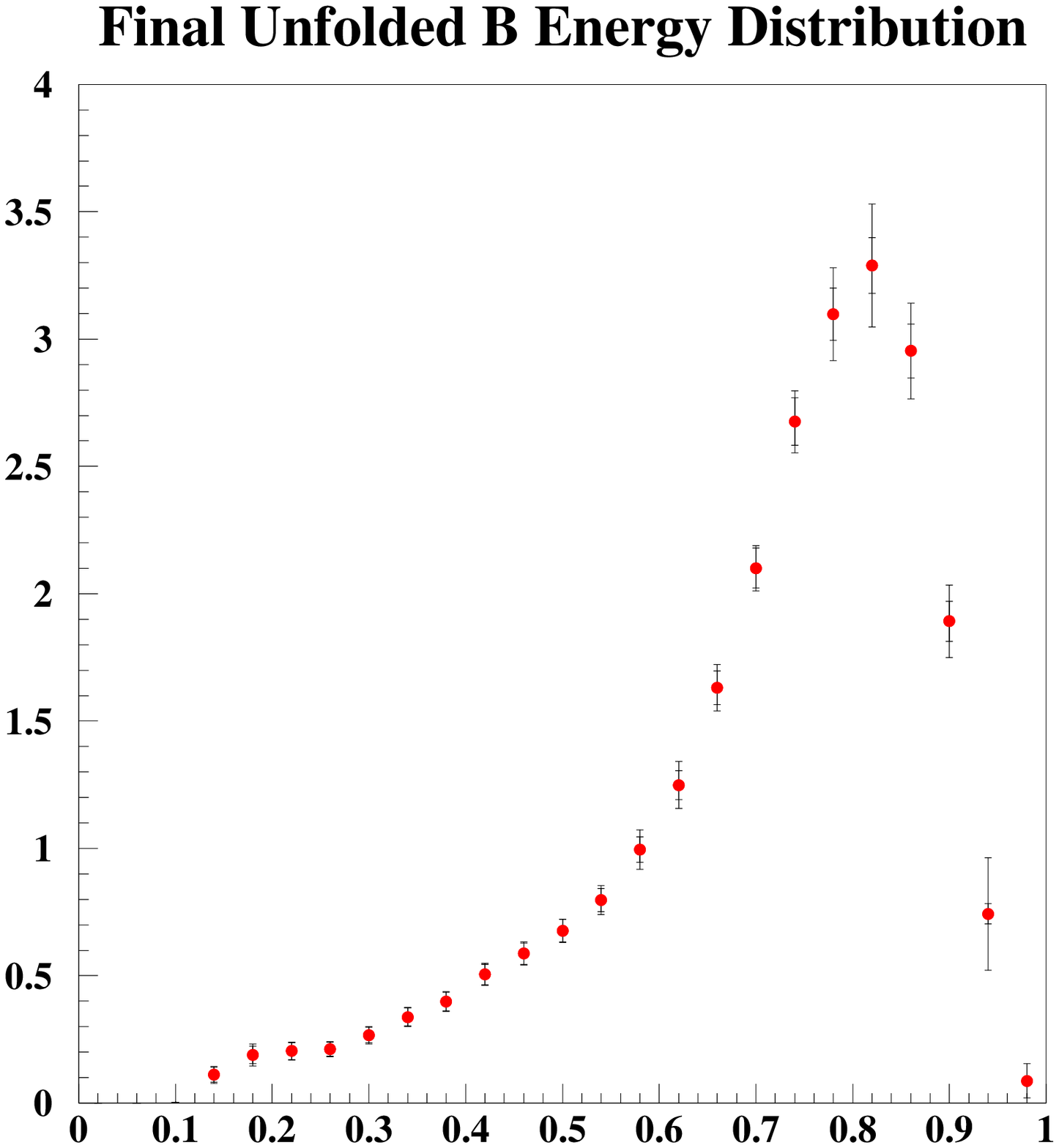}
  { \small \vspace{-0.73cm}
\begin{center}
{$\textstyle x_B$}
\end{center}
Figure 6.
The binwise average of the seven unfolded distributions (preliminary).
\vspace{0.15cm}
}
} 
\\
\indent

We then test several heavy quark fragmentation models.
Since the fragmentation functions are usually functions of an 
experimentally inaccessible variable (e.g. $z=(E+p_{\|})_{H}/(E+p_{\|})_Q$), 
it is necessary to use a Monte Carlo generator 
such as JETSET~\cite{jetset} 
to generate events according to a given input heavy
quark fragmentation function.  The resulting $B$ energy distribution is
then used to re-weigh the Monte Carlo distribution before 
comparing with the data.  
The minimum $\chi^2$ is found by varying the input parameter(s).
Within the context of the JETSET Monte Carlo, 
Kartvelishvili~\cite{kart}, Bowler~\cite{bowler}, and the Lund~\cite{lund}
models are consistent with the data, while 
BCFY~\cite{BCFY}, Collins and Spiller~\cite{collins}, and 
Peterson~\cite{pete} models are found to be inconsistent with the data.

The four functional forms and the three fragmentation models 
consistent with the data are then used in turn to 
calculate a model-dependent unfolding matrix 
which is applied to the data to 
obtain the unfolded $x_B$ distribution. The resulting seven unfolded
$x_B$ distributions show substantial model-variations.  Figure 6 shows 
the binwise average of the seven unfolded distributions,  where the inner 
error bar represents the statistical error 
and the outer error bar 
is the quadrature sum of the r.m.s.\ of the seven unfolded distributions 
and the statistical error within each bin.  
The outer error bars therefore provide an envelope within which the true $x_B$ 
distribution is expected to vary.  The 
mean of the scaled weakly decaying $B$ hadron energy distribution 
is obtained by taking the average of the means of the seven functions.
The r.m.s.\ of the seven means is regarded as the error on model-dependence.
The preliminary result is
\begin{eqnarray}
<x_B> = 0.713\pm 0.005 (stat)\pm0.007 (syst)\pm0.002 (model)
\label{eqn:average}
\end{eqnarray}
\noindent where the small model-dependence error
indicates that $<x_B>$ is relatively insensitive 
to the allowed forms of the shape of the fragmentation function.

\section*{Acknowledgements}
We would like to thank the organizers for their invitation 
and their efforts to make this conference very enjoyable.
We also thank the NSF for providing partial funding for this trip. 
We thank the personnel of the SLAC accelerator department and the
technical staffs of our collaborating institutions for their 
outstanding efforts on our behalf.  
\vskip .5truecm

\vbox{\footnotesize\renewcommand{\baselinestretch}{1}\noindent
$^*$Work supported by Department of Energy
  contracts:
  DE-FG02-91ER40676 (BU),
  DE-FG03-91ER40618 (UCSB),
  DE-FG03-92ER40689 (UCSC),
  DE-FG03-93ER40788 (CSU),
  DE-FG02-91ER40672 (Colorado),
  DE-FG02-91ER40677 (Illinois),
  DE-AC03-76SF00098 (LBL),
  DE-FG02-92ER40715 (Massachusetts),
  DE-FC02-94ER40818 (MIT),
  DE-FG03-96ER40969 (Oregon),
  DE-AC03-76SF00515 (SLAC),
  DE-FG05-91ER40627 (Tennessee),
  DE-FG02-95ER40896 (Wisconsin),
  DE-FG02-92ER40704 (Yale);
  National Science Foundation grants:
  PHY-91-13428 (UCSC),
  PHY-89-21320 (Columbia),
  PHY-92-04239 (Cincinnati),
  PHY-95-10439 (Rutgers),
  PHY-88-19316 (Vanderbilt),
  PHY-92-03212 (Washington);
  The UK Particle Physics and Astronomy Research Council
  (Brunel, Oxford and RAL);
  The Istituto Nazionale di Fisica Nucleare of Italy
  (Bologna, Ferrara, Frascati, Pisa, Padova, Perugia);
  The Japan-US Cooperative Research Project on High Energy Physics
  (Nagoya, Tohoku);
  The Korea Research Foundation (Soongsil, 1997).}

\vskip 1truecm
  
\section*{$^{**}$List of Authors} 
%
%
%
\begin{center}
\def\iADEL{$^{(1)}$}
\def\iAOMORI{$^{(2)}$}
\def\iBOLO{$^{(3)}$}
\def\iBRUN{$^{(4)}$}
\def\iBU{$^{(5)}$}
\def\iCINC{$^{(6)}$}
\def\iCOLO{$^{(7)}$}
\def\iCOLU{$^{(8)}$}
\def\iCSU{$^{(9)}$}
\def\iFERR{$^{(10)}$}
\def\iFRAS{$^{(11)}$}
\def\iILLI{$^{(12)}$}
\def\iLBL{$^{(13)}$}
\def\iLTU{$^{(14)}$}
\def\iMASS{$^{(15)}$}
\def\iMISSI{$^{(16)}$}
\def\iMIT{$^{(17)}$}
\def\iMOSCOW{$^{(18)}$}
\def\iNAGO{$^{(19)}$}
\def\iOREG{$^{(20)}$}
\def\iOXF{$^{(21)}$}
\def\iPADO{$^{(22)}$}
\def\iPERU{$^{(23)}$}
\def\iPISA{$^{(24)}$}
\def\iRAL{$^{(25)}$}
\def\iRUTG{$^{(26)}$}
\def\iSLAC{$^{(27)}$}
\def\iSOGA{$^{(28)}$}
\def\iSOONG{$^{(29)}$}
\def\iTENN{$^{(30)}$}
\def\iTOHO{$^{(31)}$}
\def\iUCSB{$^{(32)}$}
\def\iUCSC{$^{(33)}$}
\def\iVAND{$^{(34)}$}
\def\iWASH{$^{(35)}$}
\def\iWISC{$^{(36)}$}
\def\iYALE{$^{(37)}$}

  \baselineskip=.75\baselineskip  
\mbox{K. Abe\unskip,\iAOMORI}
\mbox{K.  Abe\unskip,\iNAGO}
\mbox{T. Abe\unskip,\iSLAC}
\mbox{I.Adam\unskip,\iSLAC}
\mbox{T.  Akagi\unskip,\iSLAC}
\mbox{N. J. Allen\unskip,\iBRUN}
\mbox{A. Arodzero\unskip,\iOREG}
\mbox{W.W. Ash\unskip,\iSLAC}
\mbox{D. Aston\unskip,\iSLAC}
\mbox{K.G. Baird\unskip,\iMASS}
\mbox{C. Baltay\unskip,\iYALE}
\mbox{H.R. Band\unskip,\iWISC}
\mbox{M.B. Barakat\unskip,\iLTU}
\mbox{O. Bardon\unskip,\iMIT}
\mbox{T.L. Barklow\unskip,\iSLAC}
\mbox{J.M. Bauer\unskip,\iMISSI}
\mbox{G. Bellodi\unskip,\iOXF}
\mbox{R. Ben-David\unskip,\iYALE}
\mbox{A.C. Benvenuti\unskip,\iBOLO}
\mbox{G.M. Bilei\unskip,\iPERU}
\mbox{D. Bisello\unskip,\iPADO}
\mbox{G. Blaylock\unskip,\iMASS}
\mbox{J.R. Bogart\unskip,\iSLAC}
\mbox{B. Bolen\unskip,\iMISSI}
\mbox{G.R. Bower\unskip,\iSLAC}
\mbox{J. E. Brau\unskip,\iOREG}
\mbox{M. Breidenbach\unskip,\iSLAC}
\mbox{W.M. Bugg\unskip,\iTENN}
\mbox{D. Burke\unskip,\iSLAC}
\mbox{T.H. Burnett\unskip,\iWASH}
\mbox{P.N. Burrows\unskip,\iOXF}
\mbox{A. Calcaterra\unskip,\iFRAS}
\mbox{D.O. Caldwell\unskip,\iUCSB}
\mbox{D. Calloway\unskip,\iSLAC}
\mbox{B. Camanzi\unskip,\iFERR}
\mbox{M. Carpinelli\unskip,\iPISA}
\mbox{R. Cassell\unskip,\iSLAC}
\mbox{R. Castaldi\unskip,\iPISA}
\mbox{A. Castro\unskip,\iPADO}
\mbox{M. Cavalli-Sforza\unskip,\iUCSC}
\mbox{A. Chou\unskip,\iSLAC}
\mbox{E. Church\unskip,\iWASH}
\mbox{H.O. Cohn\unskip,\iTENN}
\mbox{J.A. Coller\unskip,\iBU}
\mbox{M.R. Convery\unskip,\iSLAC}
\mbox{V. Cook\unskip,\iWASH}
\mbox{R. Cotton\unskip,\iBRUN}
\mbox{R.F. Cowan\unskip,\iMIT}
\mbox{D.G. Coyne\unskip,\iUCSC}
\mbox{G. Crawford\unskip,\iSLAC}
\mbox{C.J.S. Damerell\unskip,\iRAL}
\mbox{M. N. Danielson\unskip,\iCOLO}
\mbox{M. Daoudi\unskip,\iSLAC}
\mbox{N. de Groot\unskip,\iSLAC}
\mbox{R. Dell'Orso\unskip,\iPERU}
\mbox{P.J. Dervan\unskip,\iBRUN}
\mbox{R. de Sangro\unskip,\iFRAS}
\mbox{M. Dima\unskip,\iCSU}
\mbox{A. D'Oliveira\unskip,\iCINC}
\mbox{D.N. Dong\unskip,\iMIT}
\mbox{P.Y.C. Du\unskip,\iTENN}
\mbox{R. Dubois\unskip,\iSLAC}
\mbox{B.I. Eisenstein\unskip,\iILLI}
\mbox{V. Eschenburg\unskip,\iMISSI}
\mbox{E. Etzion\unskip,\iWISC}
\mbox{S. Fahey\unskip,\iCOLO}
\mbox{D. Falciai\unskip,\iFRAS}
\mbox{C. Fan\unskip,\iCOLO}
\mbox{J.P. Fernandez\unskip,\iUCSC}
\mbox{M.J. Fero\unskip,\iMIT}
\mbox{K.Flood\unskip,\iMASS}
\mbox{R. Frey\unskip,\iOREG}
\mbox{T. Gillman\unskip,\iRAL}
\mbox{G. Gladding\unskip,\iILLI}
\mbox{S. Gonzalez\unskip,\iMIT}
\mbox{E.L. Hart\unskip,\iTENN}
\mbox{J.L. Harton\unskip,\iCSU}
\mbox{A. Hasan\unskip,\iBRUN}
\mbox{K. Hasuko\unskip,\iTOHO}
\mbox{S. J. Hedges\unskip,\iBU}
\mbox{S.S. Hertzbach\unskip,\iMASS}
\mbox{M.D. Hildreth\unskip,\iSLAC}
\mbox{J. Huber\unskip,\iOREG}
\mbox{M.E. Huffer\unskip,\iSLAC}
\mbox{E.W. Hughes\unskip,\iSLAC}
\mbox{X.Huynh\unskip,\iSLAC}
\mbox{H. Hwang\unskip,\iOREG}
\mbox{M. Iwasaki\unskip,\iOREG}
\mbox{D. J. Jackson\unskip,\iRAL}
\mbox{P. Jacques\unskip,\iRUTG}
\mbox{J.A. Jaros\unskip,\iSLAC}
\mbox{Z.Y. Jiang\unskip,\iSLAC}
\mbox{A.S. Johnson\unskip,\iSLAC}
\mbox{J.R. Johnson\unskip,\iWISC}
\mbox{R.A. Johnson\unskip,\iCINC}
\mbox{T. Junk\unskip,\iSLAC}
\mbox{R. Kajikawa\unskip,\iNAGO}
\mbox{M. Kalelkar\unskip,\iRUTG}
\mbox{Y. Kamyshkov\unskip,\iTENN}
\mbox{H.J. Kang\unskip,\iRUTG}
\mbox{I. Karliner\unskip,\iILLI}
\mbox{H. Kawahara\unskip,\iSLAC}
\mbox{Y. D. Kim\unskip,\iSOGA}
\mbox{R. King\unskip,\iSLAC}
\mbox{M.E. King\unskip,\iSLAC}
\mbox{R.R. Kofler\unskip,\iMASS}
\mbox{N.M. Krishna\unskip,\iCOLO}
\mbox{R.S. Kroeger\unskip,\iMISSI}
\mbox{M. Langston\unskip,\iOREG}
\mbox{A. Lath\unskip,\iMIT}
\mbox{D.W.G. Leith\unskip,\iSLAC}
\mbox{V. Lia\unskip,\iMIT}
\mbox{C.-J. S. Lin\unskip,\iSLAC}
\mbox{X. Liu\unskip,\iUCSC}
\mbox{M.X. Liu\unskip,\iYALE}
\mbox{M. Loreti\unskip,\iPADO}
\mbox{A. Lu\unskip,\iUCSB}
\mbox{H.L. Lynch\unskip,\iSLAC}
\mbox{J. Ma\unskip,\iWASH}
\mbox{G. Mancinelli\unskip,\iRUTG}
\mbox{S. Manly\unskip,\iYALE}
\mbox{G. Mantovani\unskip,\iPERU}
\mbox{T.W. Markiewicz\unskip,\iSLAC}
\mbox{T. Maruyama\unskip,\iSLAC}
\mbox{H. Masuda\unskip,\iSLAC}
\mbox{E. Mazzucato\unskip,\iFERR}
\mbox{A.K. McKemey\unskip,\iBRUN}
\mbox{B.T. Meadows\unskip,\iCINC}
\mbox{G. Menegatti\unskip,\iFERR}
\mbox{R. Messner\unskip,\iSLAC}
\mbox{P.M. Mockett\unskip,\iWASH}
\mbox{K.C. Moffeit\unskip,\iSLAC}
\mbox{T.B. Moore\unskip,\iYALE}
\mbox{M.Morii\unskip,\iSLAC}
\mbox{D. Muller\unskip,\iSLAC}
\mbox{V.Murzin\unskip,\iMOSCOW}
\mbox{T. Nagamine\unskip,\iTOHO}
\mbox{S. Narita\unskip,\iTOHO}
\mbox{U. Nauenberg\unskip,\iCOLO}
\mbox{H. Neal\unskip,\iSLAC}
\mbox{M. Nussbaum\unskip,\iCINC}
\mbox{N.Oishi\unskip,\iNAGO}
\mbox{D. Onoprienko\unskip,\iTENN}
\mbox{L.S. Osborne\unskip,\iMIT}
\mbox{R.S. Panvini\unskip,\iVAND}
\mbox{H. Park\unskip,\iOREG}
\mbox{C. H. Park\unskip,\iSOONG}
\mbox{T.J. Pavel\unskip,\iSLAC}
\mbox{I. Peruzzi\unskip,\iFRAS}
\mbox{M. Piccolo\unskip,\iFRAS}
\mbox{L. Piemontese\unskip,\iFERR}
\mbox{E. Pieroni\unskip,\iPISA}
\mbox{K.T. Pitts\unskip,\iOREG}
\mbox{R.J. Plano\unskip,\iRUTG}
\mbox{R. Prepost\unskip,\iWISC}
\mbox{C.Y. Prescott\unskip,\iSLAC}
\mbox{G.D. Punkar\unskip,\iSLAC}
\mbox{J. Quigley\unskip,\iMIT}
\mbox{B.N. Ratcliff\unskip,\iSLAC}
\mbox{T.W. Reeves\unskip,\iVAND}
\mbox{J. Reidy\unskip,\iMISSI}
\mbox{P.L. Reinertsen\unskip,\iUCSC}
\mbox{P.E. Rensing\unskip,\iSLAC}
\mbox{L.S. Rochester\unskip,\iSLAC}
\mbox{P.C. Rowson\unskip,\iCOLU}
\mbox{J.J. Russell\unskip,\iSLAC}
\mbox{O.H. Saxton\unskip,\iSLAC}
\mbox{T. Schalk\unskip,\iUCSC}
\mbox{R.H. Schindler\unskip,\iSLAC}
\mbox{B.A. Schumm\unskip,\iUCSC}
\mbox{J. Schwiening\unskip,\iSLAC}
\mbox{S. Sen\unskip,\iYALE}
\mbox{V.V. Serbo\unskip,\iWISC}
\mbox{M.H. Shaevitz\unskip,\iCOLU}
\mbox{J.T. Shank\unskip,\iBU}
\mbox{G. Shapiro\unskip,\iLBL}
\mbox{D.J. Sherden\unskip,\iSLAC}
\mbox{K. D. Shmakov\unskip,\iTENN}
\mbox{C. Simopoulos\unskip,\iSLAC}
\mbox{N.B. Sinev\unskip,\iOREG}
\mbox{S.R. Smith\unskip,\iSLAC}
\mbox{M. B. Smy\unskip,\iCSU}
\mbox{J.A. Snyder\unskip,\iYALE}
\mbox{H. Staengle\unskip,\iCSU}
\mbox{A. Stahl\unskip,\iSLAC}
\mbox{P. Stamer\unskip,\iRUTG}
\mbox{R. Steiner\unskip,\iADEL}
\mbox{H. Steiner\unskip,\iLBL}
\mbox{M.G. Strauss\unskip,\iMASS}
\mbox{D. Su\unskip,\iSLAC}
\mbox{F. Suekane\unskip,\iTOHO}
\mbox{A. Sugiyama\unskip,\iNAGO}
\mbox{S. Suzuki\unskip,\iNAGO}
\mbox{M. Swartz\unskip,\iSLAC}
\mbox{A. Szumilo\unskip,\iWASH}
\mbox{T. Takahashi\unskip,\iSLAC}
\mbox{F.E. Taylor\unskip,\iMIT}
\mbox{J. Thom\unskip,\iSLAC}
\mbox{E. Torrence\unskip,\iMIT}
\mbox{N. K. Toumbas\unskip,\iSLAC}
\mbox{A.I. Trandafir\unskip,\iMASS}
\mbox{J.D. Turk\unskip,\iYALE}
\mbox{T. Usher\unskip,\iSLAC}
\mbox{C. Vannini\unskip,\iPISA}
\mbox{J. Va'vra\unskip,\iSLAC}
\mbox{E. Vella\unskip,\iSLAC}
\mbox{J.P. Venuti\unskip,\iVAND}
\mbox{R. Verdier\unskip,\iMIT}
\mbox{P.G. Verdini\unskip,\iPISA}
\mbox{S.R. Wagner\unskip,\iSLAC}
\mbox{D. L. Wagner\unskip,\iCOLO}
\mbox{A.P. Waite\unskip,\iSLAC}
\mbox{Walston, S.\unskip,\iOREG}
\mbox{J.Wang\unskip,\iSLAC}
\mbox{C. Ward\unskip,\iBRUN}
\mbox{S.J. Watts\unskip,\iBRUN}
\mbox{A.W. Weidemann\unskip,\iTENN}
\mbox{E. R. Weiss\unskip,\iWASH}
\mbox{J.S. Whitaker\unskip,\iBU}
\mbox{S.L. White\unskip,\iTENN}
\mbox{F.J. Wickens\unskip,\iRAL}
\mbox{B. Williams\unskip,\iCOLO}
\mbox{D.C. Williams\unskip,\iMIT}
\mbox{S.H. Williams\unskip,\iSLAC}
\mbox{S. Willocq\unskip,\iSLAC}
\mbox{R.J. Wilson\unskip,\iCSU}
\mbox{W.J. Wisniewski\unskip,\iSLAC}
\mbox{J. L. Wittlin\unskip,\iMASS}
\mbox{M. Woods\unskip,\iSLAC}
\mbox{G.B. Word\unskip,\iVAND}
\mbox{T.R. Wright\unskip,\iWISC}
\mbox{J. Wyss\unskip,\iPADO}
\mbox{R.K. Yamamoto\unskip,\iMIT}
\mbox{J.M. Yamartino\unskip,\iMIT}
\mbox{X. Yang\unskip,\iOREG}
\mbox{J. Yashima\unskip,\iTOHO}
\mbox{S.J. Yellin\unskip,\iUCSB}
\mbox{C.C. Young\unskip,\iSLAC}
\mbox{H. Yuta\unskip,\iAOMORI}
\mbox{G. Zapalac\unskip,\iWISC}
\mbox{R.W. Zdarko\unskip,\iSLAC}
\mbox{J. Zhou\unskip.\iOREG}

\it
  \vskip \baselineskip                   
  \centerline{(The SLD Collaboration)}   
  \vskip \baselineskip        
  \baselineskip=.75\baselineskip   
\iADEL
  Adelphi University,
  South Avenue-   Garden City,NY 11530, \break
\iAOMORI
  Aomori University,
  2-3-1 Kohata, Aomori City, 030 Japan, \break
\iBOLO
  INFN Sezione di Bologna,
  Via Irnerio 46    I-40126 Bologna  (Italy), \break
\iBRUN
  Brunel University,
  Uxbridge, Middlesex - UB8 3PH United Kingdom, \break
\iBU
  Boston University,
  590 Commonwealth Ave. - Boston,MA 02215, \break
\iCINC
  University of Cincinnati,
  Cincinnati,OH 45221, \break
\iCOLO
  University of Colorado,
  Campus Box 390 - Boulder,CO 80309, \break
\iCOLU
  Columbia University,
  Nevis Laboratories  P.O.Box 137 - Irvington,NY 10533, \break
\iCSU
  Colorado State University,
  Ft. Collins,CO 80523, \break
\iFERR
  INFN Sezione di Ferrara,
  Via Paradiso,12 - I-44100 Ferrara (Italy), \break
\iFRAS
  Lab. Nazionali di Frascati,
  Casella Postale 13   I-00044 Frascati (Italy), \break
\iILLI
  University of Illinois,
  1110 West Green St.  Urbana,IL 61801, \break
\iLBL
  Lawrence Berkeley Laboratory,
  Dept.of Physics 50B-5211 University of California-  Berkeley,CA 94720, \break
\iLTU
  Louisiana Technical University,
  , \break
\iMASS
  University of Massachusetts,
  Amherst,MA 01003, \break
\iMISSI
  University of Mississippi,
  University,MS 38677, \break
\iMIT
  Massachusetts Institute of Technology,
  77 Massachussetts Avenue  Cambridge,MA 02139, \break
\iMOSCOW
  Moscow State University,
  Institute of Nuclear Physics  119899 Moscow  Russia, \break
\iNAGO
  Nagoya University,
  Nagoya 464 Japan, \break
\iOREG
  University of Oregon,
  Department of Physics  Eugene,OR 97403, \break
\iOXF
  Oxford University,
  Oxford, OX1 3RH, United Kingdom, \break
\iPADO
  Universita di Padova,
  Via F. Marzolo,8   I-35100 Padova (Italy), \break
\iPERU
  Universita di Perugia, Sezione INFN,
  Via A. Pascoli  I-06100 Perugia (Italy), \break
\iPISA
  INFN, Sezione di Pisa,
  Via Livornese,582/AS  Piero a Grado  I-56010 Pisa (Italy), \break
\iRAL
  Rutherford Appleton Laboratory,
  Chiton,Didcot - Oxon OX11 0QX United Kingdom, \break
\iRUTG
  Rutgers University,
  Serin Physics Labs  Piscataway,NJ 08855-0849, \break
\iSLAC
  Stanford Linear Accelerator Center,
  2575 Sand Hill Road  Menlo Park,CA 94025, \break
\iSOGA
  Sogang University,
  Ricci Hall  Seoul, Korea, \break
\iSOONG
  Soongsil University,
  Dongjakgu Sangdo 5 dong 1-1    Seoul, Korea 156-743, \break
\iTENN
  University of Tennessee,
  401 A.H. Nielsen Physics Blg.  -  Knoxville,Tennessee 37996-1200, \break
\iTOHO
  Tohoku University,
  Bubble Chamber Lab. - Aramaki - Sendai 980 (Japan), \break
\iUCSB
  U.C. Santa Barbara,
  3019 Broida Hall  Santa Barbara,CA 93106, \break
\iUCSC
  U.C. Santa Cruz,
  Santa Cruz,CA 95064, \break
\iVAND
  Vanderbilt University,
  Stevenson Center,Room 5333  P.O.Box 1807,Station B  Nashville,TN 37235,
\break
\iWASH
  University of Washington,
  Seattle,WA 98105, \break
\iWISC
  University of Wisconsin,
  1150 University Avenue  Madison,WS 53706, \break
\iYALE
  Yale University,
  5th Floor Gibbs Lab. - P.O.Box 208121 - New Haven,CT 06520-8121. \break

\rm
%

\end{center}

\vskip 1truecm

\end{document}